\PassOptionsToPackage{bookmarksnumbered,unicode,colorlinks=true,linktocpage=true}{hyperref}
\documentclass[sigconf]{acmart}
\usepackage{graphicx}
\usepackage{colortbl}
\usepackage{amsmath}
\usepackage{natbib}
\usepackage{amsfonts}
\usepackage[noend]{algorithmic}
\usepackage[nameinlink,capitalize]{cleveref}
\usepackage{float}
\usepackage{algorithm}
\usepackage{subcaption}
\usepackage{dsfont}
\usepackage{bbm}
\usepackage{multirow}
\usepackage{adjustbox}
\usepackage{array}
\usepackage{enumitem}
\setlist[itemize]{leftmargin=4mm}
\usepackage[colorinlistoftodos,textsize=tiny]{todonotes} %
\usepackage{multibib}
\newcites{AP}{Appendix References}

\graphicspath{{images//}}
\DeclareGraphicsExtensions{.eps,.png,.pdf,.jpg,.tif,.tiff,.ps}

\usepackage{soul}
\usepackage{xspace}
\definecolor{navy}{rgb}{0.1, 0.1, 0.8}
\definecolor[named]{gray}{rgb}{0.4, 0.4, 0.4}
\definecolor[named]{olive}{rgb}{0.1, 0.5, 0.1}
\definecolor[named]{ruby}{rgb}{0.8, 0.1, 0.3}
\definecolor{darkpastelgreen}{rgb}{0.01, 0.75, 0.24}
\definecolor{celestialblue}{rgb}{0.29, 0.59, 0.82}
\definecolor{coral}{rgb}{1.0, 0.5, 0.31}
\definecolor{Goldenrod}{rgb}{0.8,0.8,0}

\setstcolor{red}

\newcommand{\eat}[1]{}

\newcommand{\revA}[1]{{#1}}
\newcommand{\verify}[1]{{}}

\newcommand{\NOTE}[2]{ }
\newcommand{\TODO}[2]{\xspace}
\newcommand{\nb}[1]{}
\newcommand{\note}[1]{\xspace}
\newcommand{\mar}[1]{\xspace}

\DeclareMathOperator{\Real}{\mathbb{R}}

\DeclareMathOperator{\His}{\mathcal{H}}

\def\BibTeX{{\rm B\kern-.05em{\sc i\kern-.025em b}\kern-.08emT\kern-.1667em\lower.7ex\hbox{E}\kern-.125emX}}

\newcolumntype{L}[1]{>{\raggedright\let\newline\\\arraybackslash\hspace{0pt}}m{#1}}
\newcolumntype{C}[1]{>{\centering\let\newline\\\arraybackslash\hspace{0pt}}m{#1}}
\newcolumntype{R}[1]{>{\raggedleft\let\newline\\\arraybackslash\hspace{0pt}}m{#1}}
\copyrightyear{2021}
\acmYear{2021}
\setcopyright{rightsretained}
\acmConference[WSDM '21]{Proceedings of the Fourteenth ACM International Conference on Web Search and Data Mining}{March 8--12, 2021}{Virtual Event, Israel}
\acmBooktitle{Proceedings of the Fourteenth ACM International Conference on Web Search and Data Mining (WSDM '21), March 8--12, 2021, Virtual Event, Israel}\acmDOI{10.1145/3437963.3441708}
\acmISBN{978-1-4503-8297-7/21/03}

\begin{document}
\fancyhead{} 

\newcommand{\titlename}{Evently: Modeling and Analyzing Reshare Cascades with Hawkes Processes}
\newcommand{\pf}[1]{\texttt{#1}}

\title{\titlename}

\author{Quyu Kong}
\affiliation{%
  \institution{Australian National University \&\\ UTS \& Data61, CSIRO}
  \city{Canberra}
  \country{Australia}}
\email{quyu.kong@anu.edu.au}

\author{Rohit Ram}
\affiliation{%
  \institution{University of Technology Sydney}
  \city{Sydney}
  \country{Australia}
}
\email{rohit.ram@uts.edu.au}

\author{Marian-Andrei Rizoiu}
\affiliation{%
  \institution{University of Technology Sydney \& Data61, CSIRO}
  \city{Sydney}
  \country{Australia}
}
\email{marian-andrei.rizoiu@uts.edu.au}

\begin{abstract}
Modeling online discourse dynamics is a core activity in understanding the spread of information, both offline and online, and emergent online behavior. There is currently a disconnect between the practitioners of online social media analysis --- usually social, political and communication scientists --- and the accessibility to tools capable of examining online discussions of users.
Here we present \pf{evently}, a tool for modeling online reshare cascades, and particularly retweet cascades, using self-exciting processes.
It provides a comprehensive set of functionalities for processing raw data from Twitter public APIs, modeling the temporal dynamics of processed retweet cascades and characterizing online users with a wide range of diffusion measures.
This tool is designed for researchers with a wide range of computer expertise, and it includes tutorials and detailed documentation. We illustrate the usage of \pf{evently} with an end-to-end analysis of online user behavior on a topical dataset relating to COVID-19.
We show that, by characterizing users solely based on how their content spreads online, we can disentangle influential users and online bots.
\end{abstract}

\maketitle

\section{Introduction}
The dissemination of information and opinion through social media, drives change in our societies today.
The existence of viral diffusion of information suggests that some users can exert a disproportionate influence on discourse~\citep{Cebrian2016}, and that ``\revA{malicious} actors'' can exploit misinformation campaigns causing societal divisiveness~\citep{Kim2019}. 
Consequently, there is a clear need for tools to analyze the dynamics and weaknesses of online discourse systems, and to \revA{characterize users} based on \revA{how their content diffuses online}. 

There seems to currently exist a disconnect between the practitioners of online social media analysis (who are most often social and political scientists, journalists or communication scientists) and tools facilitating this analysis.
The latter --- when they exist --- either require extensive programming experience, or make particular unrealistic assumptions about the usage flow.
The result is that practitioners carefully curate large social media datasets, which remain underutilized due to the lack of accessible tools.
This work aims to fill this gap by proposing an \revA{\pf{R} package} aimed at non-computing experts\revA{ --- admittedly featuring some quantitative expertise ---}, to analyze online discussions and users from the view of information reshare cascades.

\begin{figure*}[!tbp]
	\centering

    \includegraphics[width=0.95\textwidth]{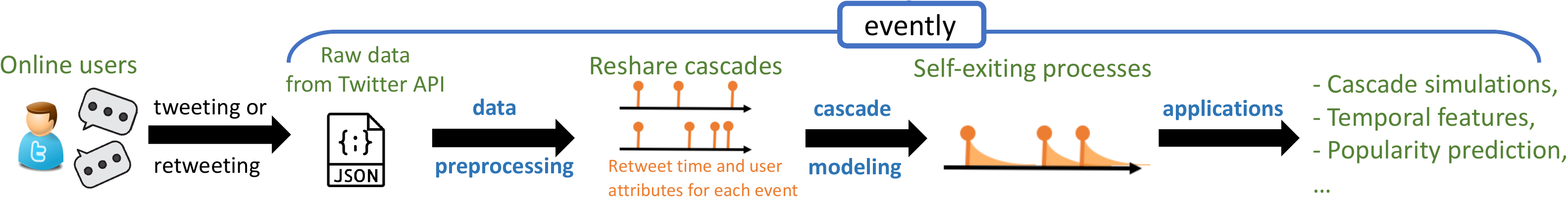}%
    \caption{
        A pipeline of functionalities (data preprocessing, cascade modeling and further applications) provided by \pf{evently} for analyzing reshare cascades of online users and characterizing the temporal dynamics of user online discussions.
    }

    \label{fig:teaser}
    \vspace{-3mm}
\end{figure*}

This work addresses two specific open questions concerning the tools \revA{designed} to model reshare cascades and analyze online users.
The first open question relates to modeling reshare cascades.
\revA{Recent works on information diffusion modeling~\citep{Zhao2015SEISMIC:Popularity,Mishra2016FeaturePrediction,Rizoiu2017c} propose only individual scripts or packages of their proposed models, often with} with disconnected API designs and (potentially complex) environmental setups.
\revA{The open} question \revA{is:} \textbf{\revA{does there exist a} tool that allows comparing multiple self-existing models on real data, while remaining easily accessible to non-experts in \revA{modeling}?}
The second open question relates to describing users based both on their activity dynamics, and how other users react to their content.
Informative temporal features of reshare cascades have been explored in prior research~\citep{martin2016exploring,Mishra2016FeaturePrediction}, but no existing tools can extract such features at the user-level.
The question is \textbf{can we extract reshare cascade features easily with a tool and show their effectiveness in online user analysis.}

In this work we address the above-mentioned open questions by introducing \pf{evently}\footnote{\label{fn:evently-source}\pf{evently} source code: \url{https://github.com/behavioral-ds/evently}}, an \pf{R} package dedicated to modeling online information reshare cascades using self-exciting point processes.
The tool is open-source and available on GitHub$^{\ref{fn:evently-source}}$, and it features extensive documentation and usage tutorials\footnote{\label{fn:evently-doc}\pf{evently} documentation and tutorials: \url{https://www.behavioral-ds.ml/evently/}}.
\addtocounter{footnote}{1}\footnotetext{\label{fn:online-tutorial}COVID-19 discussions online tutorial: \url{https://github.com/behavioral-ds/user-analysis}}
It currently supports fitting and sampling realizations of Hawkes processes~\citep{hawkes1971spectra} and variants, using several decaying kernels, both unmarked and with continuous event marks.
\pf{evently} exposes a number of functionalities around reshare cascades and online users.
For online cascades, it can fit any of its supported models to observed data, and it can sample synthetic cascades from fitted models.
It can be used to continue likely unfoldings of partially observed cascades, and compute their expected final popularity.
For online users, \pf{evently} can jointly fit all cascades initiated by the same users and obtain a descriptive model for the user.
It also allows to build a large number of dynamic user descriptors, such as the viral score (i.e., the expected size of a cascades posted by the user), and summaries of cascade sizes. Starting from a dataset containing one day-worth of Twitter discussion around COVID-19, we showcase the usage of the tool to analyze the reshare cascades and the online users.

\noindent \textbf{The main contributions of this paper are:}
\begin{itemize}
    \item \pf{evently} --- a software package dedicated to modeling reshare cascades, and capable of characterizing online users based on the reshare dynamics of the cascades they generate.
    \item A set of online tutorials showcasing how the tool can be used by non-experts, and an example analysis of discussions around COVID-19 on Twitter.
\end{itemize}

\noindent {\bf Related work.} Most prior works provide the proposed models as scripts mainly for reproducing experimental results~\citep{Mishra2016FeaturePrediction,Rizoiu2017c,kong2019modeling}.
\citet{Zhao2015SEISMIC:Popularity} ship their model in an R package as an accessible tool for predicting the final popularity of retweet cascade. 
Unlike the above which are mostly developed for demonstration purposes, \pf{evently} is designed with extendible, multi-purpose, unified set of APIs for modeling with different Hawkes process variants. 

There exist other tools that implement multiple models, which emphasize specific aspects such as language-specific implementations (e.g., \pf{THAP}~\citep{xu2017thap} in matlab, \pf{PoPPy}~\citep{xu2018poppy} in PyTorch), or network Hawkes models (\pf{pyhawkes}~\citep{linderman2014discovering}). Among these, a Python package, \pf{tick}\citep{bacry2017tick}, has the most active community and supplies a comprehensive set of models and helper functions for general time-dependent modeling including Hawkes processes.
\pf{evently} differs from these toolboxes in two ways: it is a Hawkes process toolbox in the native R language with limited dependencies; it is designed with an emphasis on online information diffusion modeling.

\section{Preliminaries}\label{sec:preliminary}
In this section, we briefly review the theoretical prerequisites concerning modeling reshare cascades using point processes.

\noindent{\bf Reshare cascades.}
\pf{evently}
analyzes the spread of online information in forms of online \textit{reshare cascades}. 
A reshare cascade consists of an initial user post and some reshare events of the post by other users. 
We denote a cascade observed up to time $T$ as $\His(T) = \{t_0, t_1, \dots\}$ where $t_i \in \His(T)$ are the event times relative to the first event ($t_0 = 0$). 
We denote cascades with additional information about events --- dubbed here as \emph{event marks} --- as
marked cascades.
We use the notation $\His_m(T) = \{(t_0, m_0), (t_1, m_1), \dots\}$, where each event is a tuple of an event time and an event mark
~\citep{Zhao2015SEISMIC:Popularity,Mishra2016FeaturePrediction}.

\noindent {\bf The Hawkes processes.} 
\pf{evently} models
reshare cascades using Hawkes processes~\citep{hawkes1971spectra} --- a type of point processes with
the self-exciting property, i.e.,
the occurrence of past events increases the likelihood of future events. 
The dynamics of event generation in a Hawkes process is controlled by its event intensity function defined as 
$\lambda(t \mid \His(T)) = \sum_{t_i < t} \phi(t - t_i)$, 
where
$\phi: \Real^+ \rightarrow \Real^+$ is a kernel function capturing the decaying influence from a historical event. 
Two widely adopted parametric forms for the kernel function $\phi$ include the exponential function $\phi_{EXP}(t) = \kappa \theta e^{-\theta t}$ and the power-law function $\phi_{PL}(t) = \kappa (t + c)^{-(1+\theta)}$.

\noindent {\bf The branching factor} $n^*$ is an important quantity for the Hawkes and HawkesN processes (as discussed in \cref{ssec:evently}) and is defined as the expected number of events directly spawned by a single event.%

\noindent {\bf The HawkesN process}~\citep{Rizoiu2017c} is a finite-population variant of the Hawkes processes. 
It assumes a finite $N$ --- the maximum number of events in the process ---,
and modulates the likelihood of future event by the remaining proportion of total population.

\noindent {\bf SEISMIC}~\citep{Zhao2015SEISMIC:Popularity} is a doubly stochastic formulation of Hawkes processes where the branching factor (dubbed as infectiousness in~\citep{Zhao2015SEISMIC:Popularity}) is a stochastic time-varying function $n^*(t)$ estimated from the observed events $\His_m(t)$.

\noindent {\bf Event simulations and parameter estimations.} We apply the rejection-sampling algorithm~\citep{ogata1981lewis} to simulate events from Hawkes and HawkesN processes and we estimate model parameters using the general log-likelihood function for point processes~\citep{Daley2008}.

\noindent{\bf Cascades joint modeling.} 
When analyzing reshare dynamics of online items (like Youtube videos and news articles) or users, it is desirable to account for multiple cascades relating to them.
\citet{kong2020exploiting} proposed to jointly model a group of cascades with a shared Hawkes model by summing the log-likelihood functions of individual cascades. 
In \cref{sec:experiment}, we model cascades initiated by same users,
and we show that the learned models can be used to separate active Twitter users from bots.

\noindent{\bf Final popularity prediction.}
The final \textit{popularity} of a reshare cascade is the total number of events which occurred until the cascade has ended. 
Predicting the final popularity of an active cascade has been extensively explored in prior works~\citep{Zhao2015SEISMIC:Popularity,Mishra2016FeaturePrediction,Rizoiu2017c}.

\noindent{\bf Viral score $v$} describes a user or an online item, and it is defined as the expected popularity of a newly started cascade relating to the given user or item.
It is obtained using the model jointly trained on all observed cascades of the user (item)~\citep{rizoiu2017online}.%

We refer to the documentation$^{\ref{fn:evently-doc}}$ for detailed mathematical definitions of aforementioned and other quantities.

\section{\pf{evently} Overview}
\label{ssec:evently}
\pf{evently} is an \pf{R} package for modeling online reshare cascades --- and retweet cascades in particular --- using Hawkes processes and their variants. 
By design, it provides an integrated set of functionalities to enable one to conduct cascade-level or user-level analysis of reshare diffusion.

\noindent {\bf Design.} \pf{evently} is designed around the interactions among three components: data (i.e., reshare cascades), models and diffusion measures. In applications, models can be used to simulate new cascades, and diffusion measures are analyzed with off-the-shelf supervised and unsupervised tools.

For cascade-level analysis, a reshare cascade is usually observed until a certain time $T$. A chosen model is then fitted on the cascade capturing its temporal dynamics. From the learned model, \pf{evently} characterizes the cascade with derived quantities such as the branching factor. It can also simulate possible future developments of the cascade after time $T$ and, in addition, derive the expectation of all future unfolding (i.e., the final popularity).

When performing user-level analysis, cascades are grouped based on the user who initiates them.
\pf{evently} models these cascades jointly, and the resulting fitted model encodes the reshare patterns at a user level. 
Similarly, new reshare cascades can be simulated from this model, and the viral score denoting the expected popularity of a new cascade from the same user can be derived. Other temporal features for the user that can be derived from the group of cascades include $6$-point summaries (mean, first/third quarters, median, minimum and maximum values) of cascade sizes, reshare event time intervals and event magnitudes~\citep{Mishra2016FeaturePrediction}.

\noindent {\bf Implementation.} \pf{evently} contains two core functions in terms of data and models: \pf{fit\_series} fits a model on given cascades; \pf{generate\_series} simulates cascades from a provided model. A model can be indicated by passing an \pf{model\_type} argument to these functions where we use abbreviated strings to denote models. For example, \textit{EXP} and \textit{PL} stands for Hawkes processes with an exponential kernel and a power-law kernel respectively, while \textit{mEXP} and \textit{mPL} are their marked variants. We refer to the package documentation$^{\ref{fn:evently-doc}}$ for a complete table of model abbreviations.

\noindent {\bf Data structure.} Cascades are structured as tables (or \pf{data.frame}s in R) where a \textit{time} column stores event timestamps relative to the first event $t_0$ and an optional \textit{magnitude} column holds the corresponding event mark information. The APIs of \pf{evently} also work with an R \pf{list} of cascade \pf{data.frame}s assuming these cascades share a same model.

\noindent {\bf Optimization.} As mentioned in~\cref{sec:preliminary}, the model parameter estimation is \revA{performed by minimizing the log-likelihood function of the point process~\citep{Daley2008}} via \pf{AMPL}, a modeling language designed to describe and solve large-scale optimization problems. Compared to other optimization tools which require precomputed or numerical gradients, \pf{AMPL} provides automatic differentiation of functions leading to model implementation efficiency. Moreover, it is also compatible with a wide range of solvers including \revA{the state-of-the-art} non-linear solver \pf{IPOPT}~\citep{Wachter2006} and \revA{the global solver} \pf{LGO}~\citep{pinter2007nonlinear}.

\noindent {\bf Installation.} 
\pf{evently} can be installed in \pf{R} directly from \textit{Github}$^{\ref{fn:evently-source}}$: \pf{remotes::install\_github(`behavioral-ds/evently')}.
It automatically configures dependencies on its first load,
which if performed manually would involve considerable effort.

\section{Case study: COVID-19 discussions}
\label{sec:experiment}
As a demonstration of \pf{evently}, we apply it to a dataset related to online discussions about the COVID-19~\citep{chen2020tracking} and present individual functionalities with code snippets and the outputs. All steps and results presented here are reproducible which can be accessed via an online \textit{Rmarkdown} notebook$^{\ref{fn:online-tutorial}}$.

\noindent {\bf Dataset.} We use a dataset of tweets concerning the novel coronavirus COVID-19 pandemic in $2020$. \citet{chen2020tracking} collected the tweet IDs via Twitter’s streaming API with a set of manually selected accounts and keywords. 
We limit our dataset to tweets posted on $31$st Jan.
The dataset is provided as a list of tweet IDs which require \textit{re-hydration} with tools like \pf{twarc}\footnote{https://github.com/DocNow/twarc}. 
As deleted tweets cannot be recovered, we obtain $68.8\%$ of the original dataset.

\noindent {\bf Import from raw data.}
Importing the COVID-19 dataset, and extracting user and cascade information can be achieved by the \pf{parse\_raw\_tweets\_to\_cascades} function from \pf{evently}.
It reads and derives pertinent information, including tweets which spawned retweet cascades during the studied period.
Our dataset contains $1,566,328$ unique tweets from $919,176$ unique users.
In total, \pf{evently} extracts $423,443$ retweet cascades, started by $280,336$ users.

\begin{figure}[!tbp]
    \centering
    \begin{subfigure}{0.44\textwidth}
        \includegraphics[width=\textwidth,page=3]{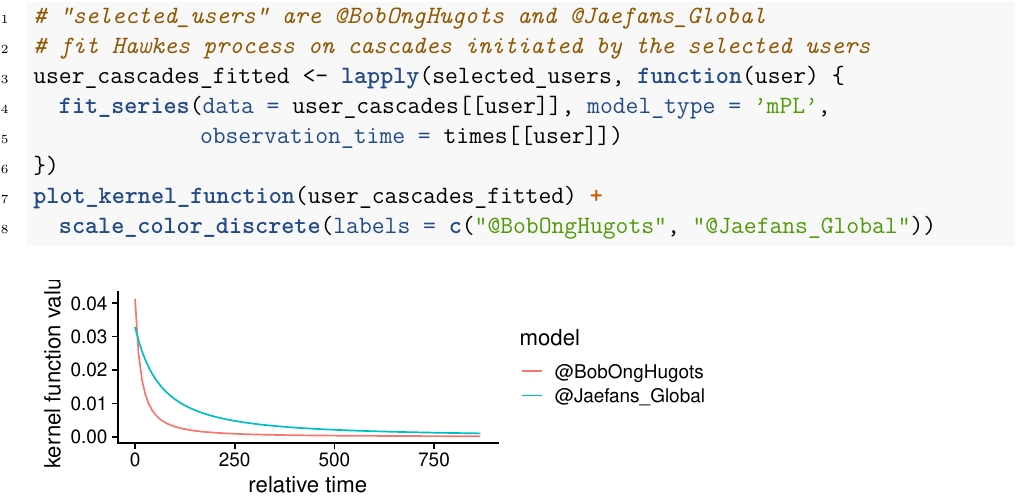}
        \vspace{-8mm}
        \caption{}
        \label{subfig:applications_d}
    \end{subfigure}
    \begin{subfigure}{0.44\textwidth}
        \includegraphics[width=\textwidth,page=4]{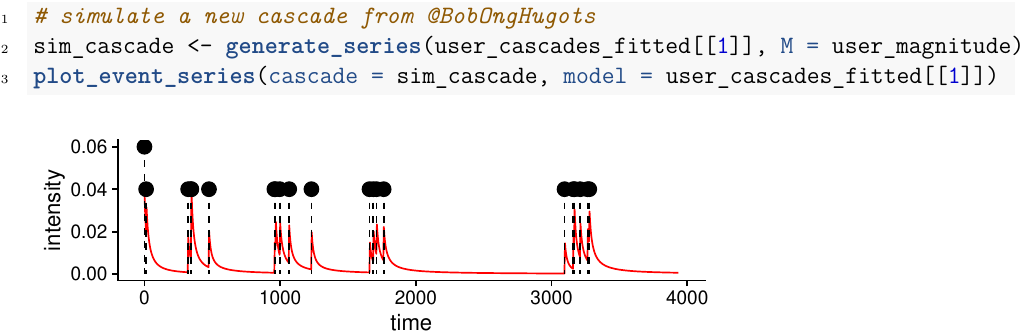}
        \vspace{-7mm}
        \caption{}
        \label{subfig:applications_e}
    \end{subfigure}
    \begin{subfigure}{0.44\textwidth}
        \includegraphics[width=\textwidth,page=5]{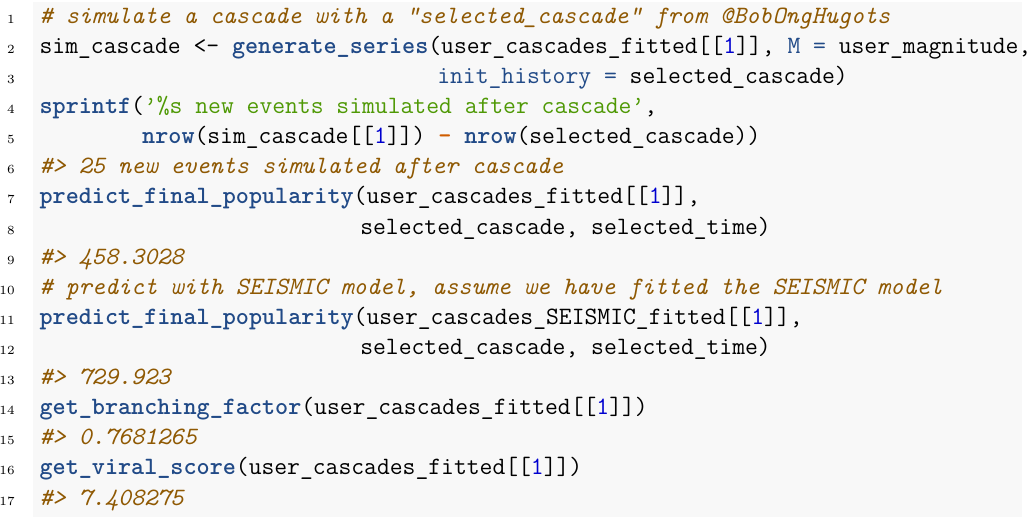}
        \vspace{-6mm}
        \caption{}
        \vspace{-4mm}
        \label{subfig:applications_f}
    \end{subfigure}
    \caption{
        Fitting and simulation of cascades from the COVID-19 dataset with \pf{evently}. Fig. (a) depicts kernel functions of learned Hawkes processes and Fig. (b) draws a simulated reshare event history with intensity values.
    }
    \label{fig:applications}
    \vspace{-6mm}
\end{figure}

\noindent {\bf Fit observed reshare cascades efficiently.} 
\pf{evently} fits Hawkes processes efficiently by leveraging the AMPL interface with a range of model choices. 
\cref{subfig:applications_d} depicts an example where we apply marked Hawkes processes with the power-law kernel function to jointly fit the cascades of 
two randomly selected Twitter users: 
\textit{@BobOngHugots} (account posting quotes from a Filipino author) and 
\textit{@Jaefans\_Global} (account of a K-pop singer), 
respectively. 
We employ the function \pf{fit\_series} from \pf{evently} and obtain the fitted models. 
The learned kernel functions for the two users are plotted at lines 7--8, and shown in the lower panel. 
We observe that, on average, tweets posted by \textit{@BobOngHugots} have an initial higher intensity but demonstrate a faster decay trend in followers' memory compared to \textit{@Jaefans\_Global}. 
On the other hand, tweets from \textit{@Jaefans\_Global} tend to influence followers for a longer period.

\noindent {\bf Simulate processes with a range of models.} 
With a given model, \pf{evently} allows to sample entire synthetic new cascades, or continue partially observed cascades. 
For instance, in line 1--3 in~\cref{subfig:applications_e} we use \pf{evently} to simulate a hypothetical cascade started by \textit{@BobOngHugots} using the model obtained in~\cref{subfig:applications_d}. 
The lower panel plots the simulated cascade which contains $21$ reshare events which is a pretty large cascade given that \textit{@BobOngHugots}'s viral score is $7.40$). 
In another example at~\cref{subfig:applications_f}, line 1--3, we partially observe a real cascade from \textit{@BobOngHugots}, and we use \pf{evently} to continue the cascade unfolding via simulation. 
$25$ new events are spawned following the observed history (line 2--6).

\noindent {\bf Compute popularity measures.} 
The above-mentioned procedure outputs just one possible ending for a given cascade. 
Using \pf{evently} we can compute the cascade's popularity, i.e. the expected cascade size over all possible unfolding.
At line 7--13, we obtain the expected final popularity with two methods: 
a marked power-law Hawkes process and 
the SEISMIC model, which output final sizes values around $458$ and $730$, respectively. We note that the true final popularity of the cascade is $472$ obtained by checking the retweets within following $10$ days ($1$st Feb to $10$th Feb).
Another two diffusion measures are computed in the example: \textit{@BobOngHugots}'s branching factor, and their viral score.

\begin{figure}[!tbp]
    \centering
        \includegraphics[width=0.35\textwidth]{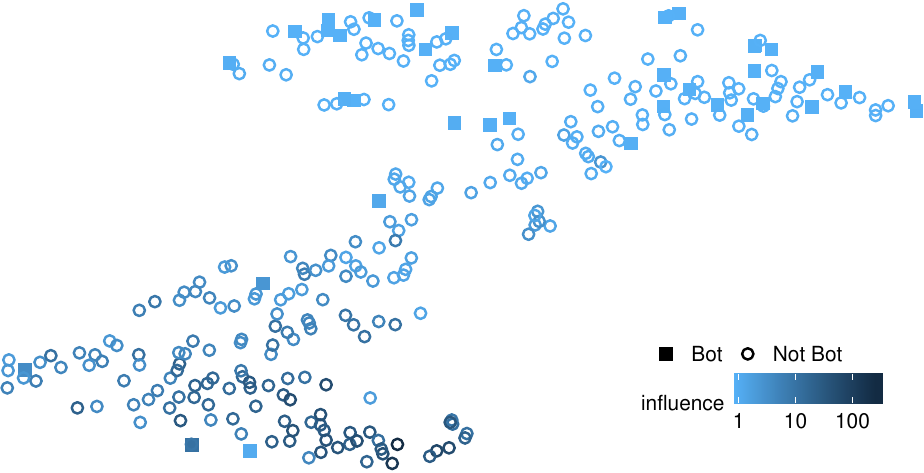}
    \caption{
        Presenting users where the positions are obtained via t-SNE~\citep{maaten2008visualizing} on temporal diffusion features from \pf{evently}. Circle colors indicate the user botness (darker blue suggests higher botness values) and circle sizes show the user influence (larger sizes mean higher influence values).
    }
    \label{fig:tsne}
   \vspace{-5mm}
\end{figure}

\noindent {\bf Visualize users in a latent space.} 
The aforementioned applications provide methods to study individual user, however it might be desirable to analyze users in relation to each other. Here we augment the user information with two additional user metrics, user \textit{influence} and \textit{botness} scores, provided by an open source tool, \pf{BirdSpotter}~\citep{rohit2021}.
We leverage the temporal features from \pf{evently} and the augmented user metrics to create a visualization of the users in the dataset. 
We select the top $300$ users who initiated the most number of cascades in the dataset. 
For these users,  we build their temporal features with \pf{evently} using the function \pf{generate\_features}. 
In~\cref{fig:tsne}, we apply the state-of-the-art dimension reduction tool t-SNE~\citep{maaten2008visualizing} to build a two-dimensional space from the higher dimensional space of the temporal features.
Finally, we label as bots the users with a \textit{botness} score higher than $0.6$~\citep{rizoiu2018debatenight}, and we color them based on
their user \textit{influence} scores.

From~\cref{fig:tsne}, we observe two obvious clusters that divide less influential users (top-right corner) from high influence users (bottom-left corner). Noticeably, most users who are classified as bots group at the top-right corner, i.e., the less influential side. On the contrary, users with high influence scores are less likely to be bots.

\section{Conclusion and Future Work}
In this work, we present, \pf{evently}, a tool for analyzing Twitter users with an emphasis on their involvement in online information diffusions.
First, we provide the theoretical background information. Then we give an overview of \pf{evently} where it models reshare cascades initiated by users. Lastly, given a dataset of tweets around COVID-19, we demonstrate the applications.

\subsection*{Acknowledgments}
\small{This research was partially funded by the National Security College, at the Australian National University through a Greenhouse Policy grant, Facebook Research under the Content Policy Research Initiative grants and the Defence Science and Technology Group of the Australian Department of Defence, through the Modelling in the Gray Zone program.}

\bibliographystyle{ACM-Reference-Format}
\bibliography{acm}

\end{document}